\def\mass#1{${\mathrm{#1\, M}_\odot}$}
\def\chem#1#2{$\mathrm{^{#2}\kern-0.8pt#1}$}
\def\reac#1#2#3#4#5#6{$\mathrm{\, ^{#2}\kern-0.8pt{#1}\, ({#3}\, ,{#4})\, {}^{#6}\kern-0.8pt{#5}\, }$}
\documentclass{aa}
\usepackage{graphics}
\usepackage{array}
\usepackage{subfigure}
\usepackage{epsfig}
\usepackage{rotating}

\begin {document}
\thesaurus{5(08.01.1; 08.06.03; 08.05.3; 08.12.1; 10.15.2 NGC 2360;
10.15.2 NGC 2447)}

\title{Chemical abundances in seven red giants of NGC 2360 and
NGC 2447\thanks{Based on observations collected at the European Southern 
Observatory, La Silla, Chile (period 56)}$^,$
\thanks{Table 3 is available only
in electronic form from the Strasbourg ftp server at 130.79.128.5}}

\author{S. Hamdani\inst{1}, P. North \inst{1}, N. Mowlavi \inst{2},
D. Raboud \inst{2}, J.-C. Mermilliod \inst{1}}

\institute{Institut d'Astronomie de l'Université de Lausanne,
           CH-1290 Chavannes-des-bois, Switzerland \and Observatoire de Genève,
           CH-1290 Sauverny, Switzerland}

\date{Received 6 March 2000/Accepted 16 June 2000}

\titlerunning{Chemical Abundances in 7 Red Giants}
\maketitle

\sloppy

\begin{abstract}
Chemical abundances of about fifteen elements from oxygen to europium are
measured in seven red giants of the two open clusters NGC 2360 and 
NGC 2447. The effective temperatures of the giants are determined
spectroscopically by taking advantage of their known masses (\mass{\sim 2}
in NGC~2360 and \mass{\sim 3} in NGC~2447) and bolometric magnitudes.

The average iron abundances we obtain for the two clusters are [Fe/H]=0.07 for
NGC~2360 and [Fe/H]=0.03 for NGC~2447.

Evolutionary stellar model calculations are performed in the mass range \mass{1-4}
in order to analyze the surface Na and O abundances predicted after the first
dredge-up. The sodium abundance shows a well defined correlation with stellar
mass in the \mass{2-3} range. The agreement between our Na abundance
determinations in NGC 2360 and our model predictions at \mass{2} is very good.
In contrast, the overabundance in one of the three stars in NGC 2447 exceeds
that predicted at \mass{3} by $\sim 0.08$~dex, which is significant
compared to the observational error bars. The effects of core overshooting,
convection prescription, metallicity and nuclear reaction rates on the
Na surface predictions of our models are investigated.

An oxygen deficiency relative to iron by 0.2~dex is measured
in our stars, in disagreement with our model predictions.
Assuming that the Sun is 0.1--0.3 dex enriched in oxygen relative to 
neighbor stars could  explain the discrepancy.

\keywords{Stars: abundances -- Stars: fundamental parameters -- Stars: evolution
-- Stars: late-type -- Open clusters and associations: individual: NGC 2360 -- 
Open clusters and associations: individual: NGC 2447} 
\end{abstract}

\section{Introduction}\label{sec1}

The comparison of chemical abundances observed at the surface of stars with those
predicted by stellar model calculations offers a valuable tool to test our
knowledge of stellar evolution. Among the most studied objects are red giants.
Those objects are formed after the main sequence (MS) phase when hydrogen has
been exhausted in the core. The core contracts, the outer layers expand, and the
convective envelope deepens into the core. The consequent mixing of the ashes
of hydrogen burning from the deep layers to the surface is called the
first dredge-up process (1DUP).

In most cases, the surface abundances after 1DUP are predicted to be
sensitive to stellar mass. The observation of red giants belonging to
given clusters presents in this respect a unique advantage since the age of the
cluster, and thus of its individual stars, is known by comparing the cluster's
distribution in the Hertzsprung-Russell diagram with predicted isochrones. The
masses of the individual stars are then determined, enabling a thorough
comparison of their surface abundances with predictions.

In this paper, we determine the abundances of oxygen, sodium and heavier
elements in four red giants of the open clusters NGC~2360 and in three red giants
of NGC~2447. All these giants belong to the clumps of their respective clusters.
The initial masses of our stars are estimated to 2 and \mass{3} for those
in NGC~2360 and NGC~2447, respectively. These masses turn out to be in a
critical range for surface sodium enhancement after the 1DUP.
Oxygen, on the other hand, is an interesting element to
study since it has already been observed to be slightly deficient relative to
iron not only in supergiants but also in giants of globular clusters (Brown \&
Wallerstein \cite{B91}) and open clusters (Luck \cite{L94}).
Finally, the abundances of the iron-group elements are used to directly
derive the metallicity of the two clusters.
This direct technique, which contrasts with the classical, but
less reliable, technique based on color indices and
photometric calibrations, has so far been used only in a small number of
open clusters (about 20, see Strobel \cite{S91}, Twarog et al. \cite{T97}).
Our observations enable to enlarge that list.

The observational material is presented in Sect.~\ref{Sect:Observations} and the
abundance determination procedure in Sect.~\ref{Sect:reduction}. The metallicity
is derived in Sect.~\ref{Sect:metallicity}, and the sodium and oxygen abundances
analyzed in Sects.~\ref{Sect:sodium} and \ref{Sect:oxygen}, respectively.
Conclusions are drawn in Sect.~\ref{Sect:conclusions}.

\section{Observational Material}
\label{Sect:Observations}

Table~\ref{Tab:clusters} summarizes some parameters characterizing the two
open clusters NGC~2360 and NGC~2447 studied in this paper.
The distance modulus has been re-determined using
Mermilliod's database (\cite{M95}, \cite{Me99}) and the age is determined
with the theoretical isochrones of Schaller et al. (\cite{S92}). They
lead to turn-off masses of 1.98 and \mass{2.75} for NGC~2360 and NGC~2447,
respectively.

\begin{table}[ht]
\caption{Adopted parameters for NGC 2360 and 2447, including the iron abundance
[Fe/H] and metallicity [$M$/H] obtained in this paper
(Sect.~\ref{Sect:metallicity}). The mass M$_{\rm turn-off}$ at the turn-off
(defined as point T in Fig. 1 of Maeder \& Meynet \cite{MM91}) is given too.}
\label{Tab:clusters}
\begin{tabular}{ccc}
\hline \\[-4pt]
&NGC 2360&NGC 2447\\
\hline \\[-4pt]
$\alpha$ (1950) & $7^{\rm h} 15.5^{\rm m}$ & $7^{\rm h} 42.5^{\rm m}$\\
$\delta$ (1950) & $-15^\circ 32'$ & $-15^\circ 32'$ \\
$m_{\rm V}-M_{\rm V}$ & 10.40 & 10.25 \\
$E(B-V)$              & 0.07  &  0.04 \\
$A_{\rm V}$           & 0.22  &  0.13 \\
Distance [kpc]        & 1.086 &  1.057 \\
Log Age               & 9.06  &  8.65 \\
M$_{\rm turn-off}$ [M$_\odot$]& 1.98  &  2.75 \\
\multicolumn{1}{c}{[Fe/H]}& 0.07  &  0.03 \\
\multicolumn{1}{c}{[$M$/H]}& 0.10  &  0.05 \\
\hline
\end{tabular}
\end{table}

 The seven red giants studied in this paper have been selected on the basis
of their constant
radial velocity, as determined by CORAVEL observations,
in order to ensure that they are single stars. This point is important since
the light of a secondary component could bias the equivalent widths used to
estimate the abundances of the primary.

For each star, one echelle spectrum has been obtained by DR in November
1995 using the EMMI spectrograph attached to the NTT 3.5m telescope at
the European Southern Observatory in La Silla, Chile. The grism \#5 and the echelle
grating \#10 were used, yielding spectra in the wavelength range from
4050\AA\ to 6650\AA\ with a resolving power of $R=28\,000$.
The detector was the ESO CCD \#36, with $2048\times 2048$ pixels.
The signal-to-noise ratio per pixel of the extracted 
spectra varies from 54 to 226, depending on the wavelength range and 
exposure time.

The journal of the observations is given in Table~\ref{Tab:journal}, together with the main
characteristics of each stars (identification, $V$ magnitude, Geneva $[B-V]$
color index, S/N ratio, Julian date, exposure time, radial velocity derived 
from our spectra and average RV from CORAVEL observations).
The radial velocities are determined from our spectra 
by cross-correlation with a CORAVEL-type binary mask optimized for the K0 
spectral type. They are compatible with those determined earlier from CORAVEL 
observations by Mermilliod \& Mayor (\cite{MM89}, \cite{MM90}), confirming that 
the red giants of our sample are very probably single.

 Observations of two IAU RV standards, HD 66141 and HD 80170
(of which Udry et al. \cite{U99} give the precise RV value from CORAVEL 
observations), show that our results can be trusted to within
$\pm 1.0$~km\,s$^{-1}$. To reach this accuracy, we had to monitor the instrumental drift by taking a few short Th-Ar calibration exposures during the 
night and correlate them with a longer calibration exposure taken in the 
preceding afternoon; then we linearly interpolated the instrumental velocity 
shifts for the epochs of the science exposures.

\begin{center}
\begin{table*}[ht]
\caption{Observational data of the 7 red giants and signal-to-noise ratio per 
pixel of their extracted spectra. The $V$ magnitude and $[B-V]$ index are those
of Geneva photometry. The adopted numbering is that of Becker et al.
(\cite{B76}), but we also give the numbers in the system of Eggen (\cite{E68})
for NGC 2360. There is no other numbering system for NGC 2447, but the map of
this cluster is available on the web (Mermilliod \cite{Me99}).The radial
velocities of the 7 red giants and of two reference stars are also given,
as determined in this work (``EMMI'') at the indicated Julian dates. The
exposure times are also listed.
The CORAVEL RV values (``COR.'') of the cluster stars are average ones from
Mermilliod \& Mayor (\cite{MM89}, \cite{MM90}), while those of the reference
stars (IAU standards) are taken from Udry et al. (\cite{U99}). The errors
quoted for the CORAVEL observations are related to the average RV values; the
last column (``N'') gives the number of CORAVEL measurements.}
\label{Tab:journal}
\begin{tabular}{cccccccrrrc}
\hline \\[-4pt]
\multicolumn{1}{c}{Star}&\multicolumn{1}{c}{DM or}&\multicolumn{1}{c}{V}
&\multicolumn{1}{c}{[B-V]}&\multicolumn{2}{c}{S/N ratio around}& JD 
& $t_{exp}$& \multicolumn{2}{c}{V$_{\rm r}$ [km\,s$^{-1}$]}& N\\
&Eggen No& &&  5150\AA\ &  6600\AA\ &-2450000&  [s]    &EMMI&COR.&COR. \\
\hline
2360-7 &   8    &11.087&0.294& 54& 89& 30.741&1800&28.53&27.2$\pm 0.2$& 5 \\
2360-50&  67    &11.082&0.311& 71&108& 30.767&2700&27.65&27.2$\pm 0.3$& 4 \\
2360-62&  81    &11.272&0.251& 73&146& 31.730&3600&28.05&27.3$\pm 0.5$& 4 \\
2360-86& 110    &10.787&0.309&114&181& 31.822&3600&26.72&27.3$\pm 0.3$& 4 \\
\hline				                                         
2447-28&-23$^\circ$ 6102& 9.849&0.226&112&170& 30.847&1800&19.98&21.2$\pm 0.2$& 3 \\
2447-34&        &10.123&0.197&143&236& 32.755&3000&23.30&22.1$\pm 0.2$& 3 \\
2447-41&        &10.031&0.204&124&226& 32.840&2700&21.40&21.5$\pm 0.2$& 3 \\
\hline
HD 66141& HR 3145& 4.406&0.637&   &   & 30.833&  90&71.43&71.6$\pm 0.3$& \\
        &        &      &     &   &   & 33.827&  90&73.10&     \\      
HD 80170& HR 3694& 5.310&0.510&   &   & 31.799& 180& 1.38& 0.5$\pm 0.2$& \\
\hline	    
\end{tabular}
\end{table*}	      
\end{center}

\section{Abundance determination}
\label{Sect:reduction}

For the reduction of the spectra, we used the Inter-Tacos software of
Geneva Observatory\footnote{The Inter-Tacos software is the same as that used for the
Elodie and Coralie spectrographs.}. The spectra were divided
by an average of six flat-field exposures, after the background had been fitted 
by 2-d polynomials and subtracted. Then, they were extracted
using a simple addition of the intensities along a virtual slit of 15 pixels.
The wavelength calibration was performed with the same software, using Th-Ar spectra
taken in the afternoon preceding the night. The rms scatter of the residuals
around the fitted dispersion relation was 3.1 m\AA, which is quite good for
this resolution. After rebinning to a constant wavelength step of 0.03~\AA,
which is over-sampled, we filtered all spectra by eliminating the high
frequencies of their Fourier Transform. For that purpose, we used a FFT with a
window of 65\% of $\Delta \lambda$ ($\Delta \lambda=\lambda$/R).
The 55 orders were individually normalized to the continuum,
using an interactive Supermongo procedure which interpolates between the 
continuum points by a 3rd degree spline.

\subsection{Abundance determination}

The list of lines and oscillator strengths used for our abundance
determinations is the same as that adopted by Boyarchuk et al. (\cite{B96}).
The photospheric lines are chosen in wavelength zones avoiding the telluric 
absorption lines. The equivalent widths
below 10 m\AA\ are rejected because they are too sensitive to noise and
normalization. Table~3 (available only in electronic form) summarizes the line
parameters and measured equivalent widths.
For Mg\,\textsc{i}, not included in the list of Boyarchuk et al., we
adopt the $\log gf$ values of Th\'evenin (\cite{T89}, \cite{T90}).

The abundances are determined using the MOOG code (C. Sneden, Texas University)
in the ``abfind'' mode, combined with atmospheric models of
Kurucz (\cite{K96}). The Kurucz models are interpolated by cubic splines in the
$T_{\rm eff}$ and $\log g$ plane (solar abundances are assumed as far as the
atmospheric structure is concerned).
For some lines for which we have reasons to suspect that blends would make
the equivalent width method risky (oxygen and europium), we
compute a synthetic spectrum including a large number of neighboring lines,
and determine the abundance of the considered element by best fitting the
synthetic spectra to the observed one (the `best' fit being evaluated by eye
estimate). In that case, the MOOG code is used in its ``synth''
mode and the lines are taken from the VALD database (Piskunov et al. \cite{P95}).

The measurement of abundance deviations as small as 0.15~dex requires good 
spectra with signal-to-noise (S/N) ratios above 100. Such a S/N is reached at 
almost all wavelengths in all our spectra, except near the [O\,\textsc{i}] line 
in three stars in NGC~2360 (star numbers 7, 50 and 62).
No O abundance is presented for those stars.

\subsection{Atmospheric parameters}
\label{Sect:atmosphere}

The determination of the atmospheric parameters
$\log g$ (gravity), $T_{\rm eff}$ (effective temperature) and
$V_{\rm t}$ (microturbulent velocity) is crucial for a
good abundance estimation. The fact that the distance to the open clusters and
the masses $M$ of the individual stars are known provides a relation between $\log g$
and $T_{eff}$ given by
\begin{equation}
\label{Eq:logg}
\log g = -12.51 + \log M + 4 \log T_{\rm eff} + 0.4 (M_{\rm V}+BC),
\end{equation}
where $M$ is expressed in solar mass, $M_{\rm V}$ is the absolute visual magnitude
in V and BC is the bolometric correction. We assume $M_{\rm bol\odot}=4.75$, and
take the bolometric corrections from Flower (\cite{F77}) (which are very close to
the theoretically derived ones of Bessell et al. \cite{B98}).

The atmospheric parameters $T_{\rm eff}$ and $V_{\rm t}$ of each star are
estimated from an iterative procedure based on the method used by
Boyarchuk et al. (\cite{B96}), but modified to take into account the known 
bolometric magnitudes of the stars.

\begin{figure}
\begin{center}
\epsfig{figure=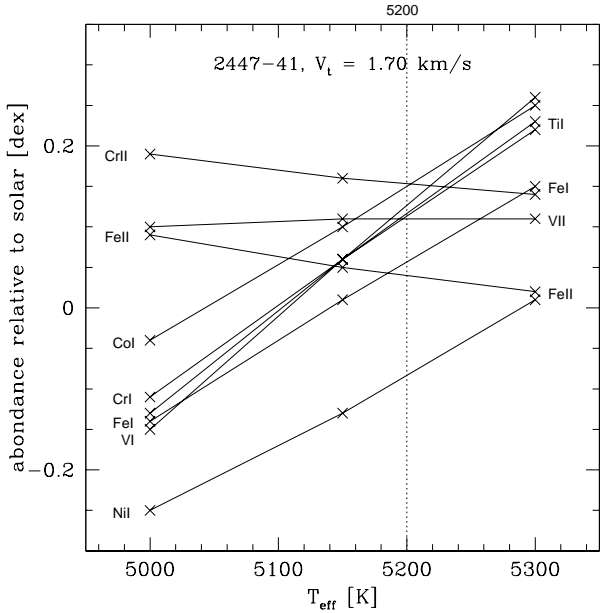,width=8.5cm}
\caption{\label{Fig:Teff}
         Abundances of the iron-group elements as a function of effective
         temperature for the star NGC~2447-41.
         The value of $\log g$ is given by Eq.~\ref{Eq:logg}
         at each effective temperature.
         The vertical line shows the adopted $T_{\rm eff}$ value.
        }
\end{center}
\end{figure}

Let us consider a given star.
Starting with an initial guess of $V_{\rm t}=1.6$~km\,s$^{-1}$, we construct
a diagram (Fig.~\ref{Fig:Teff}) plotting the abundances of the iron-peak
elements computed by MOOG from our observed spectrum, as a function of the 
effective temperature ($\log g$ is computed for each $T_{\rm eff}$ value 
through Eq. 1). If all iron-peak elements had
a solar abundance distribution, then there should be a value of $T_{\rm eff}$
at which their abundances (normalized to solar) are identical.
In practice, we choose (by visual estimate) the value of $T_{\rm eff}$
at which the spread of abundances is minimal. 
Then, for this $T_{\rm eff}$ value, the usual diagram showing the Fe\,\textsc{i}
abundances versus equivalent widths is used to adjust the best velocity of
microturbulence $V_{\rm t}$. Finally, the whole procedure is reiterated in
order to get mutually consistent parameters.

The value of $V_{\rm t}$ found after the convergence of the iterative procedure
differs by no more than 0.15 km\,s$^{-1}$ with the first guess, which is
consistent with the formal standard deviation of 0.11~km\,s$^{-1}$
for that parameter, obtained from the uncertainty on the regression line of
Fe\,\textsc{i} abundances versus equivalent widths.

This iterative procedure is applied to each of our stars. The resulting
atmospheric parameters are summarized in Table~\ref{Tab:abundances}.
The derived effective temperatures agree to better than 100~K with the classical
one based on the lack of correlation between abundance and 
excitation potential, the agreement improving with S/N ratio.

Finally, the sensitivity of our abundance determinations to the adopted
atmosphere models is tested by recomputing some abundances with the models of
Bell et al. (\cite{BEG76}). They lead to similar results, in the sense that
abundances given by individual lines are the same within $\pm 0.07$~dex after
allowance for a slight systematic offset: the models of Bell et al.
systematically lead to a 0.04~dex underabundance compared to
Kurucz's models (for lines with an equivalent width $W_\lambda < 100$~m\AA; the
offset is larger for stronger lines).
In view of the relatively good agreement between both sets of models, we chose
the Kurucz ones because they are more recent and easier to use.

The surface abundance of a given element is determined by adjusting the
equivalent width(s) of its line(s) predicted by synthetic spectra to those
measured in the observed spectra. Actually the MOOG program computes an 
abundance for each line of a given element, and derives a mean abundance value 
by averaging the abundances associated to each of those lines. In computing the
synthetic spectra, the formation of a dozen molecules (among them C$_2$, CH, CN, CO, MgH, TiO) is considered.

In the case of Eu, the only usable line is blended. We therefore
determine its abundance by a visual match on a synthetic spectrum which includes
all possible blends. This element is marked with an asterisk
in Table~3. The same technique is used for oxygen, discussed in
Sect.~\ref{Sect:oxygen}. The $\lambda 6645.1$ Eu\,\textsc{ii} line does not seem to be significantly affected by hyperfine structure, since its equivalent 
width in the Sun provides an abundance perfectly consistent with the meteoritic 
one, using precise experimental oscillator strengths (Bi\'emont et al. 
\cite{BK82}). Our $\log gf$ value adopted for this line is only 0.03 dex below 
the one used by Bi\'emont et al., so it would lead to a solar Eu abundance of 
0.53 instead of 0.50 dex, while the meteoritic abundance is $0.54\pm 0.01$~dex
(Grevesse \& Noels \cite{GN93}).

The elemental abundances found in this work are listed in Table~\ref{Tab:abundances}.

\begin{figure}
\begin{center}
\epsfig{figure=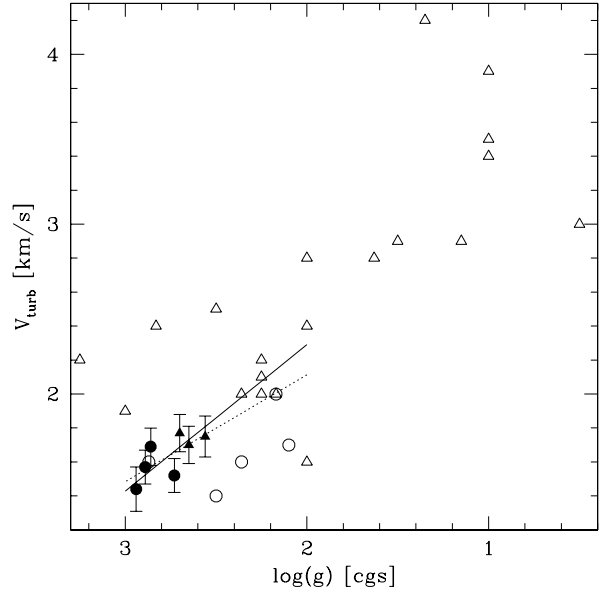,width=8.5cm}
\caption{\label{Fig:Vt}
         Microturbulent velocity as a function of surface gravity.
         Full dots: NGC 2360; full triangles: NGC 2447; open dots: field giants
         of Boyarchuk et al. (\cite{B96}); open triangles: cluster giants and
         supergiants of Luck (\cite{L94}) with $4500\leq T_{\rm eff} \leq 5500$~K.
         Full line: regression line fitted to our seven stars assuming the same
         error on both axes;
         dotted line: regression line assuming no error on the abscissa.
        }
\end{center}
\end{figure}

\subsection{$V_{\rm t}$ versus $\log g$}
For information, we plot in Fig.~\ref{Fig:Vt} (filled circles and triangles)
the microturbulence
velocity deduced for our stars as a function of the surface gravity, together
with the measurements of Boyarchuk et al. (\cite{B96}) for five G7-K0 field giants (open circles, their sixth star is not considered because it is much 
cooler, of M0 type) and those of Luck (\cite{L94}) for clusters' supergiants and giants with $4500\leq T_{\rm eff} \leq 5500$~K (open triangles). The combined data reveals a positive correlation between $V_{\rm t}$ and $\log g$.

Considering only our seven points, the usual correlation coefficient
$\rho = -0.697$ gives a $t$-test value of only $-2.17$, implying a significance 
level of about 90 percent, while the Spearman rank correlation coefficient 
$\rho_{\rm Spearman} = -0.786$ gives a slightly better $t$-test of $-2.84$ and a significance level of 96 percent. For the usual LSQ fit assuming no error on the abscissa and a constant error on the ordinate, we find
\begin{equation}
V_{\rm t}=(-0.63\pm 0.29)\log g + (3.37\pm 0.80)
\end{equation}
with an rms scatter of the residuals of 0.083~km\,s$^{-1}$.
For a fit where errors are assumed to be the same on both axes, the relation
would be
\begin{equation}
V_{\rm t}= -0.86 \log g + 4.01
\end{equation}
with an rms scatter of the residuals of 0.088~km\,s$^{-1}$. The scatter of
the residuals is entirely compatible with the internal standard deviation
of $V_{\rm t}$ given above. It is smaller than that of the data from Luck
(\cite{L94}) and from Boyarchuk et al. (\cite{B96}), partly due to the
better homogeneity in metallicity and effective temperature.

\begin{figure*}
\begin{center}
\begin{minipage}[t]{.46\linewidth}
\epsfig{figure=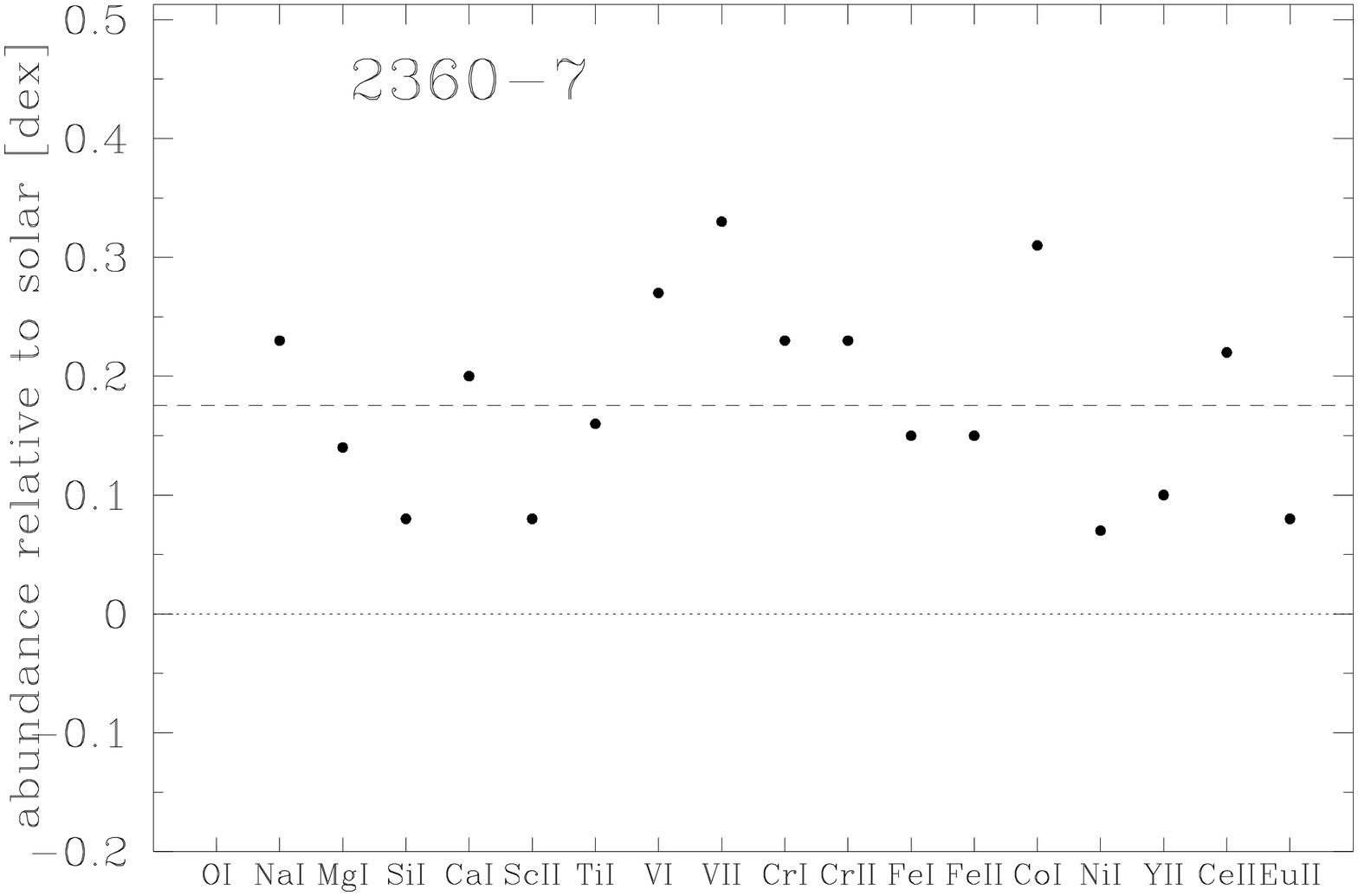,width=7.5cm}
\end{minipage}
\begin{minipage}[t]{.46\linewidth}
\epsfig{figure=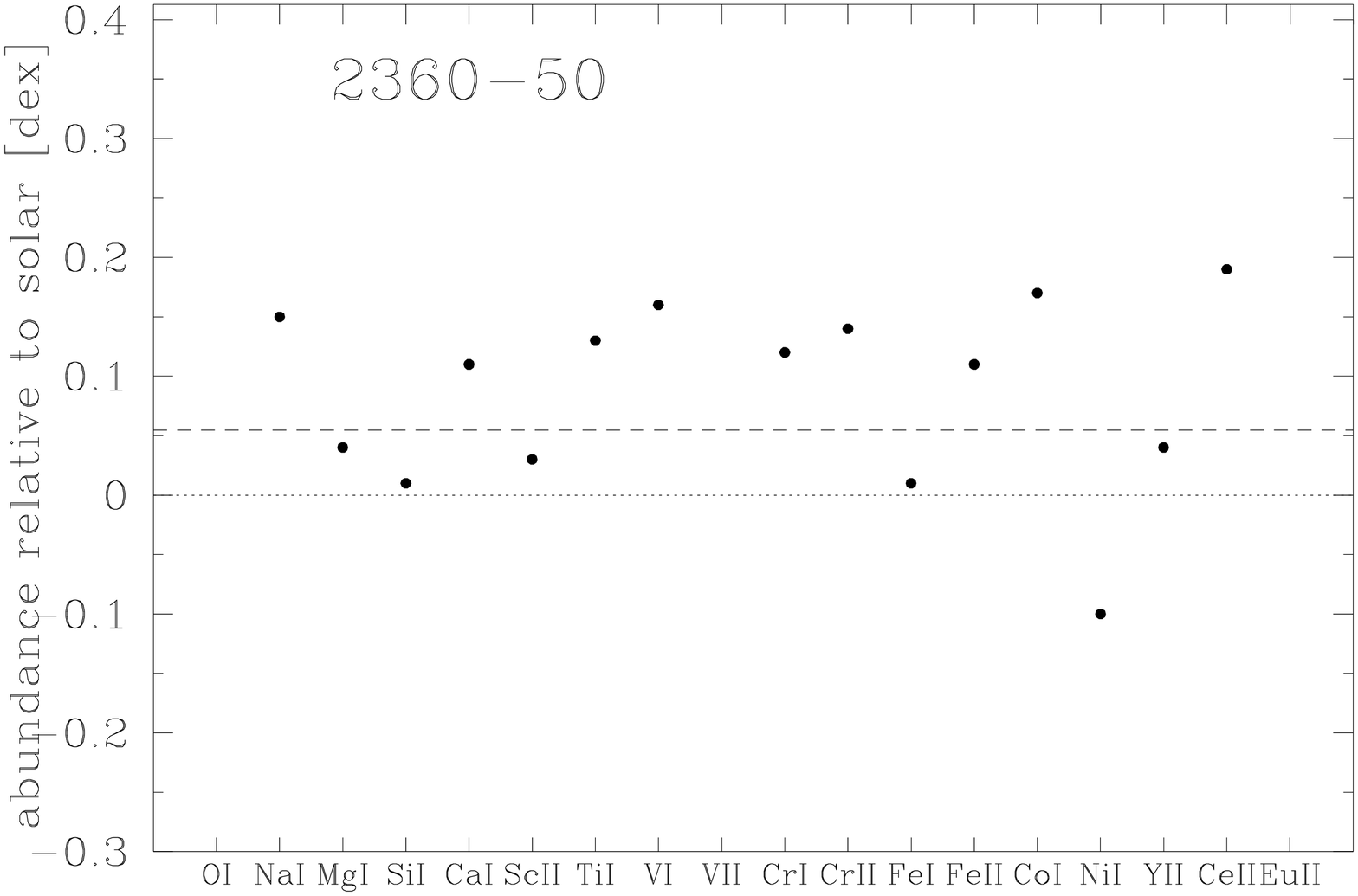,width=7.5cm}
\end{minipage}
\begin{minipage}[ht]{.46\linewidth}
\epsfig{figure=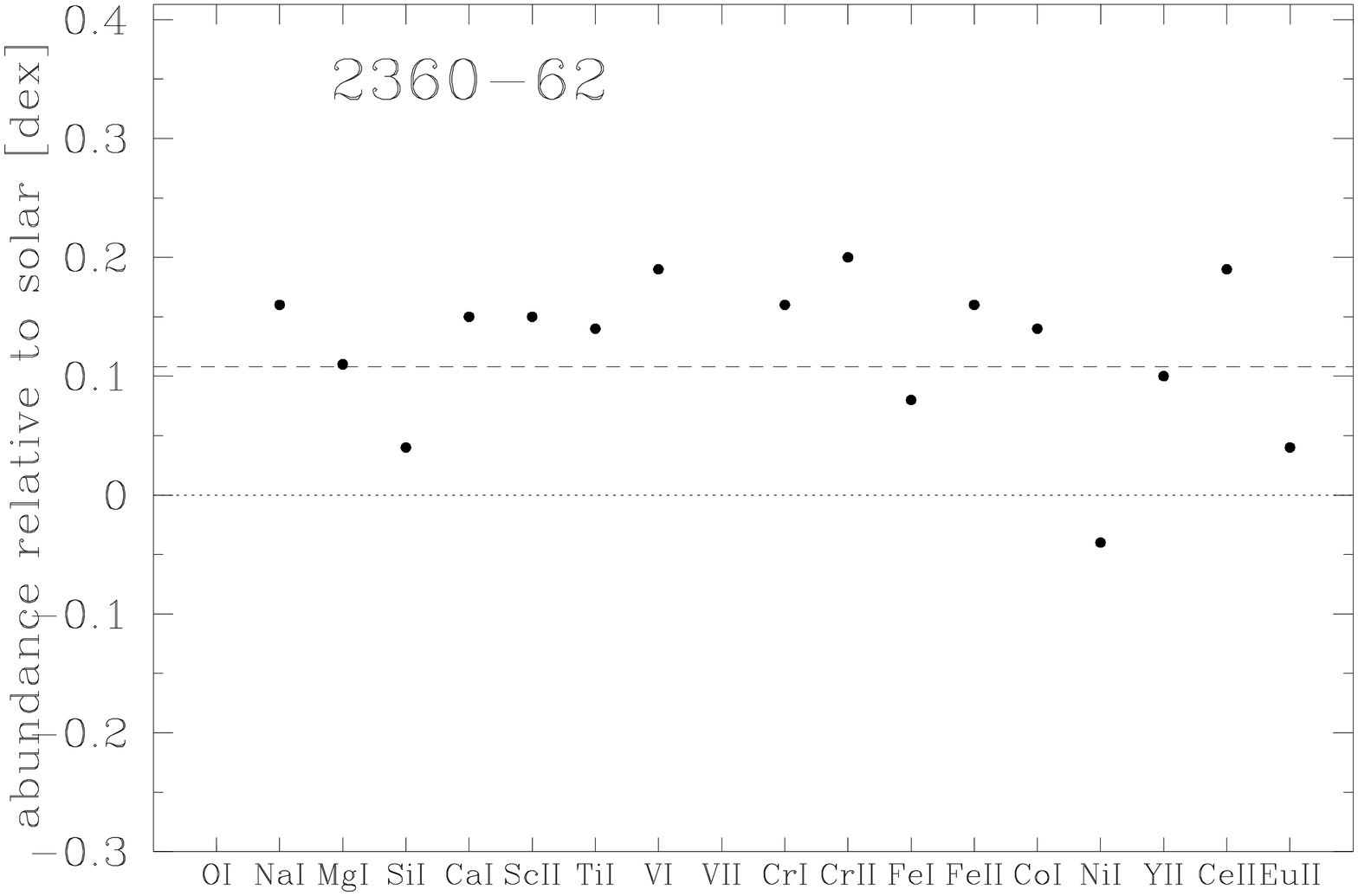,width=7.5cm}
\end{minipage}
\begin{minipage}[ht]{.46\linewidth}
\epsfig{figure=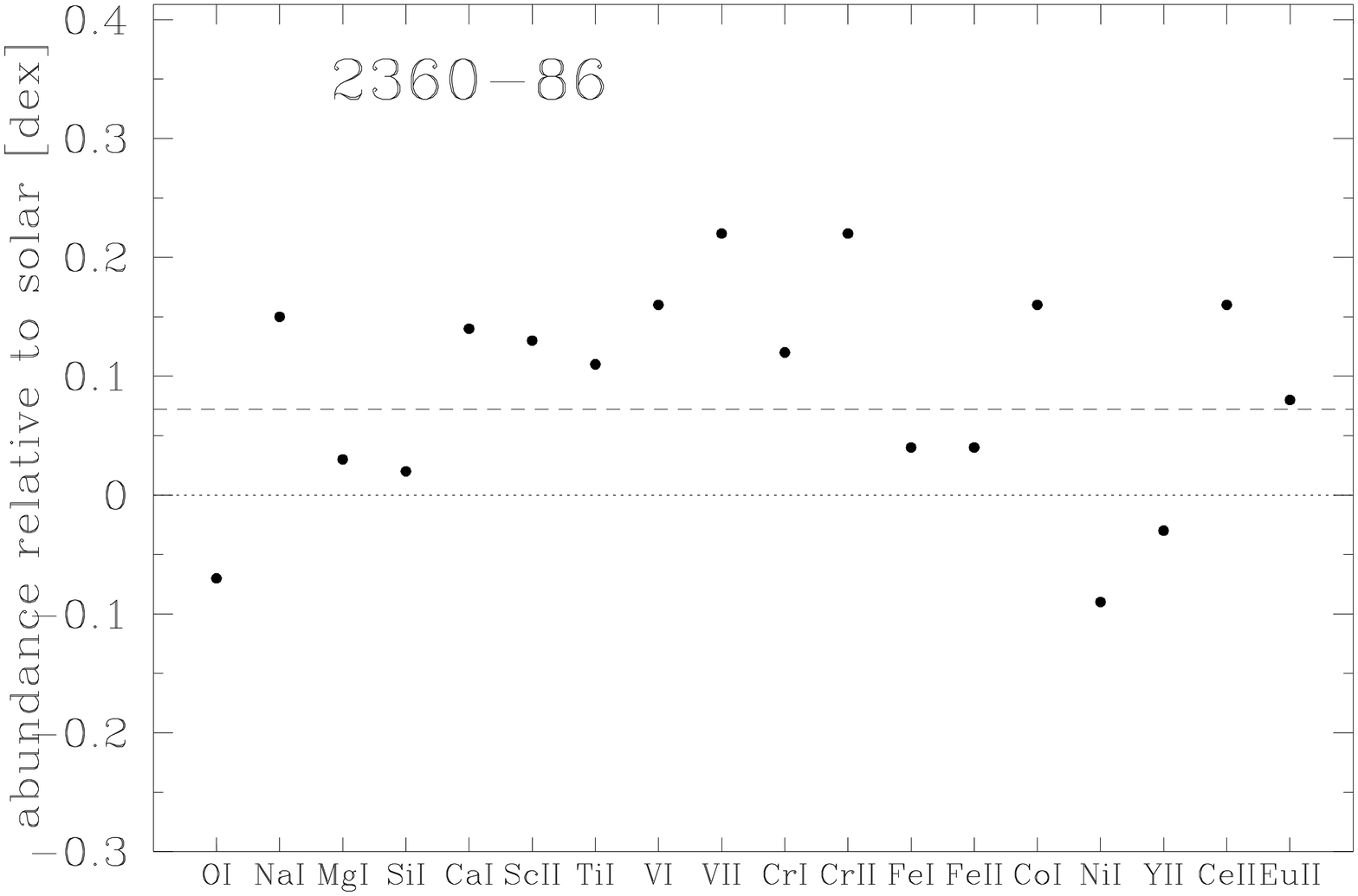,width=7.5cm}
\end{minipage}
\begin{minipage}[ht]{.46\linewidth}
\epsfig{figure=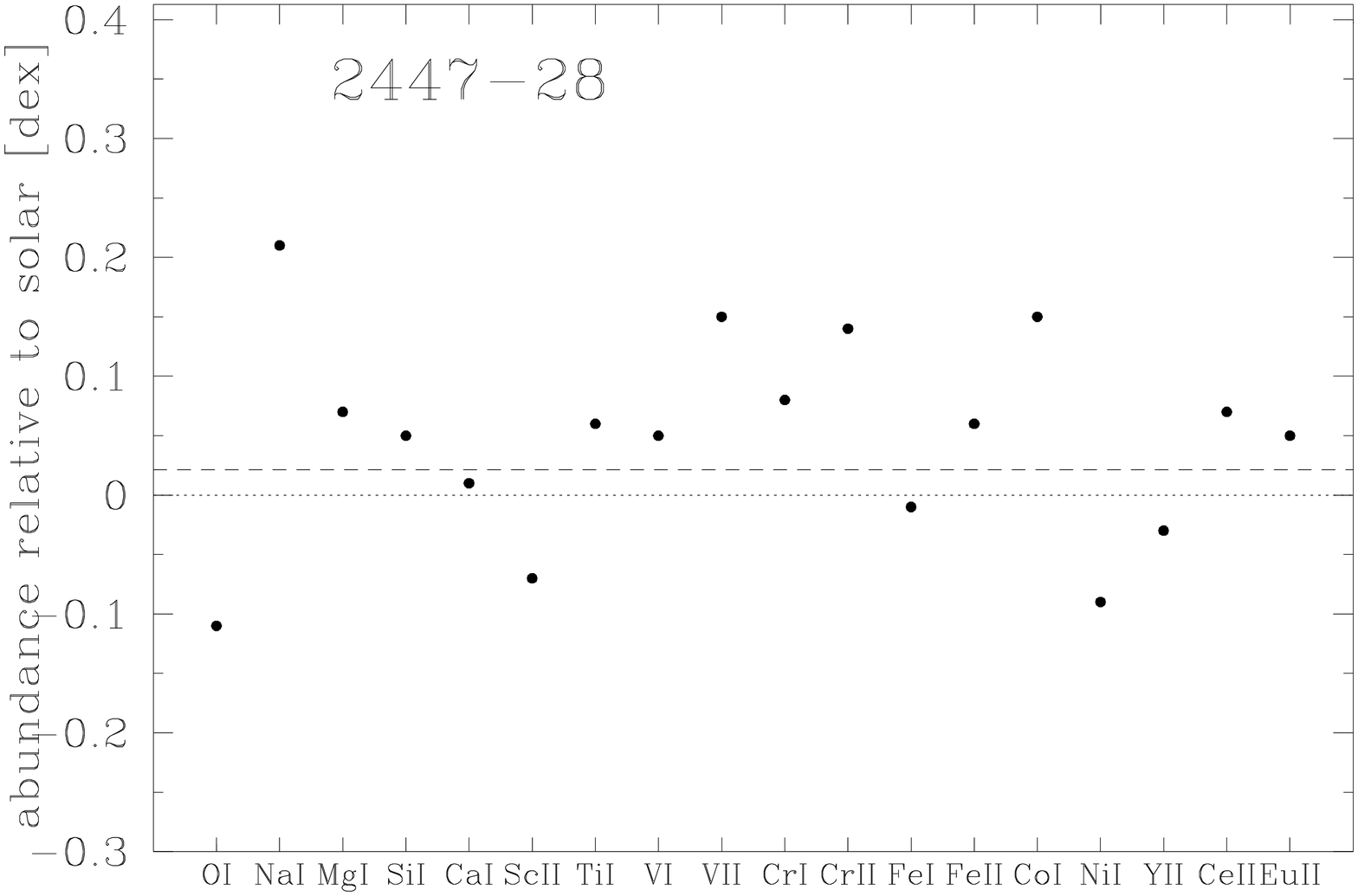,width=7.5cm}
\end{minipage}
\begin{minipage}[ht]{.46\linewidth}
\epsfig{figure=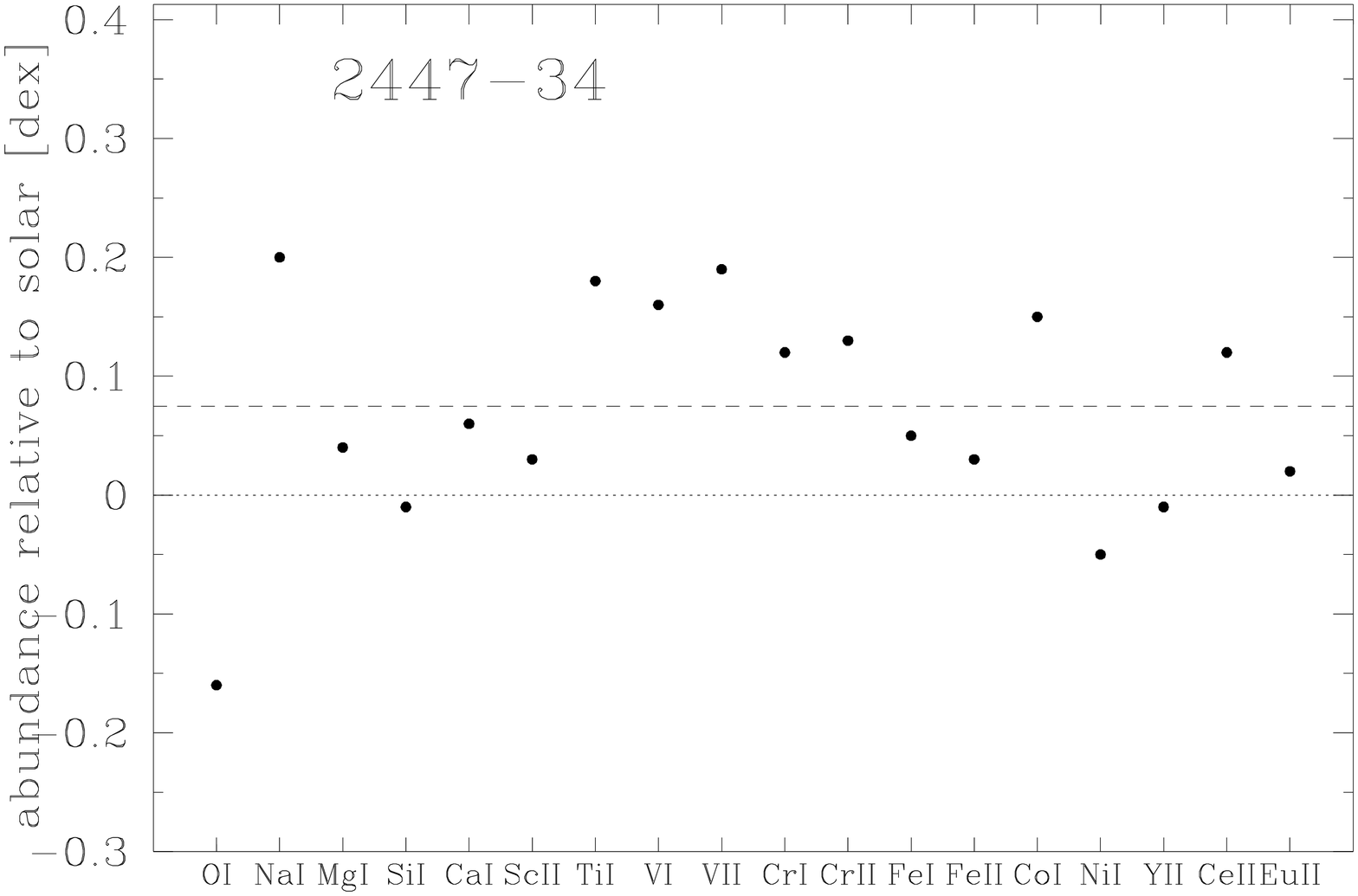,width=7.5cm}
\end{minipage}
\begin{minipage}[ht]{.46\linewidth}
\epsfig{file=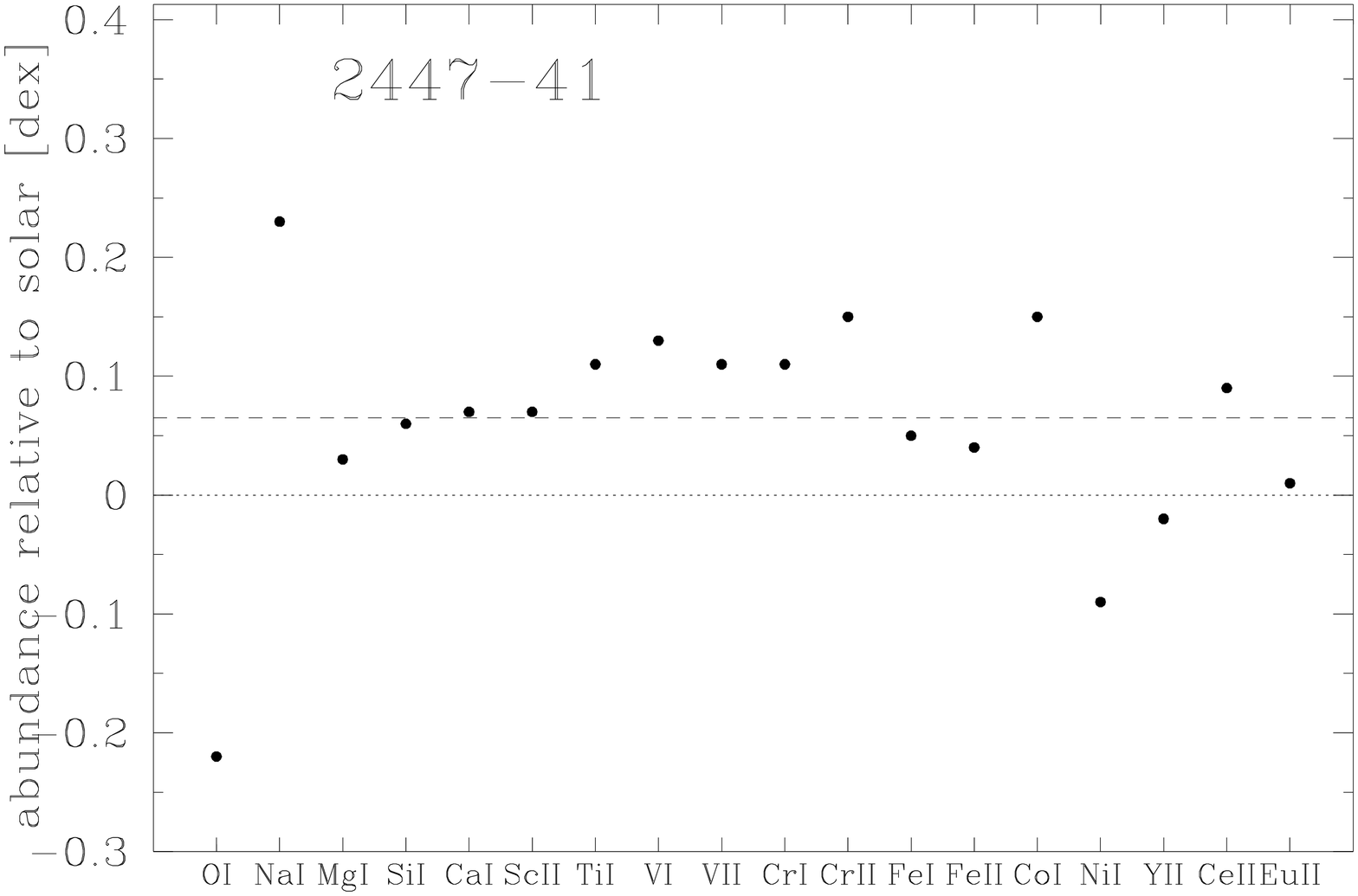,width=7.5cm}
\end{minipage}
\caption{Abundances relative to solar for the 7 red giants. The dotted line
represents the solar abundances, the dashed line is [$M$/H], defined as
the average abundance of the iron-peak elements weighted by the number of 
lines.}
\label{Fig:abundances}
\end{center}
\end{figure*}

\section{Metallicity}
\label{Sect:metallicity}
It is difficult to derive an error estimate on the abundances.
If we assume that each star has a solar distribution of heavy elements,
then an error estimate is given by the dispersion of the heavy element
abundances around the mean abundance (relative to solar). The abundances
of all measured elements are shown in Fig.~\ref{Fig:abundances} for each of our 
stars, with the mean iron-peak element abundance indicated by a dashed line in 
each panel. From visual inspection (and excluding Ni which seems
underabundant in some of our stars, probably only because of some 
systematic error in the oscillator strengths), we estimate the error on the 
abundance of a given element to be $\pm 0.1$~dex. Table~\ref{Tab:abundances} 
summarizes the abundances derived in all our stars
together with the number of lines used for their determination.

\setcounter{table}{3}
\begin{table*}
\begin{center}
\caption{Fundamental parameters of the 7 red giants and mean abundances 
relative to the Sun. For each species, the second line gives the rms
standard deviation of the abundances obtained from individual lines.
N is the number of lines measured.
The oxygen abundances and [O/Fe] are derived from the
equivalent width neglecting the possible contribution of the
Ni\,\textsc{i} $\lambda 6300.33$ line, while [O/Fe]$^*$
is computed from a synthetic spectrum including the Ni\,\textsc{i} $\lambda 6300.33$
line (with the $gf$ value from Kurucz \cite{K94}). [Fe/H] is an average weighted by the number of Fe\,\textsc{i} and Fe\,\textsc{ii} lines. [$M$/H] is a
similar average, but on all iron-peak elements from Ca to Ni.}
\label{Tab:abundances}
\def\cg{[}
\def\cd{]}
\begin{tabular}{|c|r|r|r|r|r|r|r|r|r|r|r|r|r|r|}
\hline
&\multicolumn{2}{|c|}{2360-7}&\multicolumn{2}{|c|}{2360-50}
&\multicolumn{2}{|c|}{2360-62}&\multicolumn{2}{|c|}{2360-86}
&\multicolumn{2}{|c|}{2447-28}&\multicolumn{2}{|c|}{2447-34}
&\multicolumn{2}{|c|}{2447-41}\\
\hline
Mass [M$_\odot$]&\multicolumn{2}{|c|}{2.10}&\multicolumn{2}{|c|}{2.10}
&\multicolumn{2}{|c|}{2.08}&\multicolumn{2}{|c|}{2.13}&\multicolumn{2}{|c|}{2.93}
&\multicolumn{2}{|c|}{2.81}&\multicolumn{2}{|c|}{2.87}\\
M$_{\rm bol}$&\multicolumn{2}{|c|}{0.51}&\multicolumn{2}{|c|}{0.49}
&\multicolumn{2}{|c|}{0.68}&\multicolumn{2}{|c|}{0.18}&\multicolumn{2}{|c|}{-0.60}
&\multicolumn{2}{|c|}{-0.30}&\multicolumn{2}{|c|}{-0.40}\\
M$_{\rm V}$&\multicolumn{2}{|c|}{0.69}&\multicolumn{2}{|c|}{0.68}
&\multicolumn{2}{|c|}{0.87}&\multicolumn{2}{|c|}{0.39}&\multicolumn{2}{|c|}{-0.40}
&\multicolumn{2}{|c|}{-0.13}&\multicolumn{2}{|c|}{-0.22}\\
$T_{\rm eff}$&\multicolumn{2}{|c|}{5230}&\multicolumn{2}{|c|}{5170}
&\multicolumn{2}{|c|}{5180}&\multicolumn{2}{|c|}{5130}&\multicolumn{2}{|c|}{5140}
&\multicolumn{2}{|c|}{5250}&\multicolumn{2}{|c|}{5200}\\
$\log g$&\multicolumn{2}{|c|}{2.89}&\multicolumn{2}{|c|}{2.86}
&\multicolumn{2}{|c|}{2.94}&\multicolumn{2}{|c|}{2.73}&\multicolumn{2}{|c|}{2.56}
&\multicolumn{2}{|c|}{2.70}&\multicolumn{2}{|c|}{2.65}\\
$V_{\rm t}$&\multicolumn{2}{|c|}{1.57}&\multicolumn{2}{|c|}{1.69}
&\multicolumn{2}{|c|}{1.44}&\multicolumn{2}{|c|}{1.52}&\multicolumn{2}{|c|}{1.75}
&\multicolumn{2}{|c|}{1.77}&\multicolumn{2}{|c|}{1.70}\\
\hline
&N&[El./H]&N&[El./H]&N&[El./H]&N&[El./H]&N&[El./H]&N&[El./H]&N&[El./H]\\ \hline
O\,\textsc{i}  &  &     &  &     &  &     & 1&-0.07& 1&-0.11& 1&-0.16& 1&-0.22\\
Na\,\textsc{i} & 2&0.23& 2&0.15& 2&0.16& 2&0.15& 2&0.21& 2&0.20& 2&0.23\\
               &  & 0.01&  & 0.04&  & 0.00&  & 0.03&  & 0.00&  & 0.00&  & 0.01\\
Mg\,\textsc{i} & 2& 0.14& 2& 0.04& 2& 0.11& 2& 0.03& 2& 0.07& 2& 0.04& 2& 0.03\\
               &  & 0.00&  & 0.04&  & 0.03&  & 0.06&  & 0.03&  & 0.08&  & 0.06\\
Si\,\textsc{i} & 7& 0.08& 7& 0.01& 7& 0.04& 7& 0.02& 6& 0.05& 7&-0.01& 7& 0.06\\
               &  & 0.09&  & 0.09&  & 0.07&  & 0.06&  & 0.13&  & 0.09&  & 0.09\\
Ca\,\textsc{i} & 9& 0.20& 9& 0.11& 9& 0.15& 9& 0.14& 9& 0.01& 9& 0.06& 9& 0.07\\
               &  & 0.11&  & 0.12&  & 0.08&  & 0.08&  & 0.16&  & 0.13&  & 0.11\\
Sc\,\textsc{ii}& 2& 0.08& 1& 0.03& 2& 0.15& 2& 0.13& 2&-0.07& 2& 0.03& 2& 0.07\\
               &  & 0.05&  &     &  & 0.09&  & 0.07&  & 0.16&  & 0.02&  & 0.03\\
Ti\,\textsc{i} & 8& 0.16& 9& 0.13& 9& 0.14& 9& 0.11& 8& 0.06& 9& 0.18& 9& 0.11\\
               &  & 0.13&  & 0.16&  & 0.17&  & 0.16&  & 0.18&  & 0.16&  & 0.25\\
V\,\textsc{i}  & 7& 0.27& 6& 0.16& 7& 0.19& 5& 0.16& 6& 0.05& 6& 0.16& 6& 0.13\\
               &  & 0.16&  & 0.11&  & 0.16&  & 0.09&  & 0.05&  & 0.10&  & 0.16\\
V\,\textsc{ii} & 1& 0.33&  &     &  &     & 1& 0.22& 2& 0.15& 1& 0.19& 2& 0.11\\
               &  &     &  &     &  &     &  &     &  & 0.07&  &     &  & 0.08\\
Cr\,\textsc{i} &18& 0.23&17& 0.12&16& 0.16&17& 0.12&16& 0.08&16& 0.12&16& 0.11\\
               &  & 0.20&  & 0.20&  & 0.18&  & 0.16&  & 0.17&  & 0.15&  & 0.18\\
Cr\,\textsc{ii}& 4& 0.23& 3& 0.14& 4& 0.20& 4& 0.22& 3& 0.14& 3& 0.13& 3& 0.15\\
               &  & 0.04&  & 0.06&  & 0.07&  & 0.10&  & 0.20&  & 0.15&  & 0.11\\
Fe\,\textsc{i} &57& 0.15&55& 0.01&52& 0.08&55& 0.04&55&-0.01&56& 0.05&57& 0.05\\
               &  & 0.17&  & 0.17&  & 0.17&  & 0.16&  & 0.21&  & 0.19&  & 0.19\\
Fe\,\textsc{ii}& 4& 0.15& 4& 0.11& 4& 0.16& 4& 0.04& 4& 0.06& 4& 0.03& 4& 0.04\\
               &  & 0.10&  & 0.20&  & 0.17&  & 0.13&  & 0.16&  & 0.13&  & 0.16\\
Co\,\textsc{i} & 6& 0.31& 7& 0.17& 7& 0.14& 7& 0.16& 8& 0.15& 9& 0.15& 9& 0.15\\
               &  & 0.28&  & 0.28&  & 0.24&  & 0.30&  & 0.25&  & 0.17&  & 0.24\\
Ni\,\textsc{i} &11& 0.07&11&-0.10&10&-0.04&11&-0.09&11&-0.09&11&-0.05&11&-0.09\\
               &  & 0.08&  & 0.16&  & 0.12&  & 0.18&  & 0.20&  & 0.18&  & 0.19\\
Y\,\textsc{ii} & 2& 0.10& 2& 0.04& 2& 0.10& 2&-0.03& 2&-0.03& 2&-0.01& 2&-0.02\\
               &  & 0.09&  & 0.08&  & 0.03&  & 0.08&  & 0.03&  & 0.11&  & 0.06\\
Ce\,\textsc{ii}& 2& 0.22& 2& 0.19& 2& 0.19& 2& 0.16& 2& 0.07& 2& 0.12& 2& 0.09\\
               &  & 0.04&  & 0.12&  & 0.14&  & 0.24&  & 0.00&  & 0.02&  & 0.01\\
Eu\,\textsc{ii}& 1& 0.08&  &     & 1& 0.04& 1& 0.08& 1& 0.05& 1& 0.02& 1& 0.01\\
 \hline
\cg Na/Fe \cd  &  & 0.08&  & 0.13&  & 0.07&  & 0.11&  & 0.22&  & 0.15&  & 0.18\\
\cg O/Fe \cd        &  &     &  &     &  &           &  &-0.11&  &-0.12&  &-0.21&  &-0.27\\
\cg O/Fe \cd $^*$   &  &     &  &     &  &$\sim -0.6$&  &-0.72&  &-0.54&  &-0.60&  &-0.72\\
 \hline
\cg Fe/H \cd   &  & 0.15&  & 0.02&  & 0.09&  & 0.04&  &-0.01&  & 0.05&  & 0.05\\
\cg $M$/H \cd  &  & 0.18&  & 0.06&  & 0.11&  & 0.07&  & 0.02&  & 0.08&  & 0.07\\
\hline
\end{tabular}
\end{center}
\end{table*}

The metallicity of each star is defined as a weighted average of 
the abundances of the iron-group elements, the weight being the
number of lines. Since these weights are roughly the same for all stars,
each cluster's metallicity is defined as the arithmetic mean of the
metallicities of its stars. The values found are
[$M$/H]=0.10 for NGC 2360 and [$M$/H]=0.05 for NGC 2447. The iron
abundances are slightly smaller: [Fe/H]=0.07 for NGC 2360 and [$M$/H]=0.03
for NGC 2447. Those are reported in Table~\ref{Tab:abundances}.

\section{Sodium}
\label{Sect:sodium}

\subsection{Abundances}
\label{Sect:Naabundances}

\begin{figure}
\center
\epsfig{file=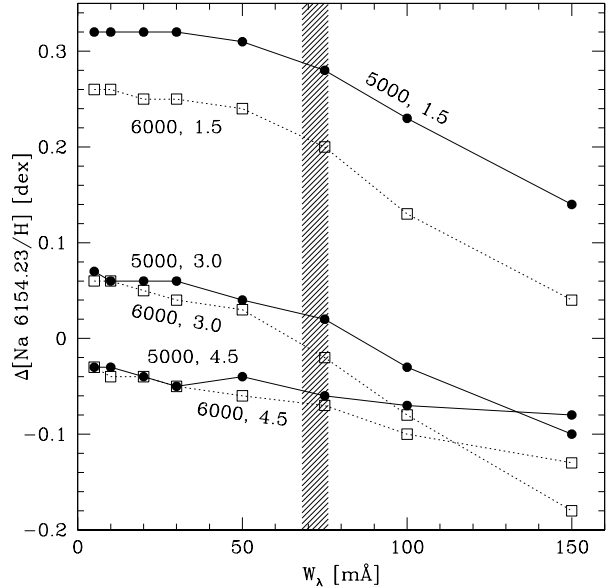,width=8.5cm}
\caption{\label{Fig:NLTE}
         NLTE effects on the Na\,\textsc{i}~$\lambda 6154.22$ line according
         to Gratton et al. (\cite{GC99}). Full dots linked with continuous curves are for
         models with $T_{\rm eff}=5000$~K and the three surface gravities
         $\log g = 1.5, 3.0, 4.5$. Open squares linked with dotted lines:
         $T_{\rm eff}=6000$~K and same $\log g$ values. The metallicity is solar
         ([$M$/H]=0.0) in all cases. The shaded area defines the range of
         equivalent widths observed in our sample of seven giants.
}
\end{figure}

The sodium overabundances [Na/Fe] derived from our spectra are summarized in
Table~\ref{Tab:abundances} for our seven stars.
They are small in NGC~2360, the oldest of the two
clusters, with [Na/Fe]= 0.07 -- 0.13 dex, and amount to 0.15 -- 0.22 dex
in NGC~2447.

The error in the Na overabundance [Na/Fe] due to the uncertainties on
$T_{\rm eff}$, $\log g$ and $V_{\rm t}$ are not large because Na\,\textsc{i}
and Fe\,\textsc{i} have similar behaviors with respect to those parameters
(e.g. Table~9 of Luck \cite{L94})
such that their effect on the [Na/Fe] ratio cancels out.
The errors on the [Na/Fe] ratio due to the continuum placement, on the one
hand, and to the photon noise, on the other hand, are estimated in the
following way. For each star, the equivalent widths of the two Na\,\textsc{i}
lines are measured five times, each time after having renormalized the spectrum
of the relevant region. The standard deviations of the equivalent widths are then
computed from these data and propagated into standard deviations of
abundances using the MOOG code. The resulting dispersion on the
Na abundance amounts to 0.02--0.05 dex, depending on the S/N ratio.
A synthetic spectrum of a typical red giant (with $T_{\rm eff}=5200$~K,
$\log g=2.77$, [$M$/H]=0.0 but [Na/Fe]=+0.15) is then produced, and a
Gaussian noise added to it for four representative S/N values, five spectra
being produced independently with the same S/N ratio. The equivalent widths of
both Na lines are measured (by a Gaussian fit) for these 20 spectra and their
standard deviation computed for each S/N ratio. The deviations are found to
lie between 2 and 3.3 percent for S/N ratios between 230 and 90. They are
translated into abundance errors (after dividing them by $\sqrt{2}$ since there
are two lines) and added quadratically to the errors due to the continuum
position. Finally, we assume that the error on the continuum position for
Fe\,\textsc{i} is similar to that for Na\,\textsc{i} and, admitting that they
are independent of each other,
add them quadratically to the total error on the Na abundance\footnote{the error
on the Fe abundance due to photon noise is neglected because of the large 
number of lines of this element} (in fact there is probably a correlation
between the continuum placement for the Fe lines and that for the Na lines,
but neglecting it only results in an overestimate of the error, so that
we stay on the safe side). The resulting estimated errors amount to about
0.03--0.07 dex, the lowest ones pertaining to NGC 2447.

Among the systematic errors which may affect the Na and Fe abundances, some
are negligible for the [Na/Fe] abundance ratio because of their mutual cancelation, as mentioned above.
This is not the case, however, for the oscillator strengths adopted in the
synthetic spectra. If the $\log gf$ values of the two Na\,\textsc{i} lines are 
slightly in error while those of Fe\,\textsc{i} are statistically correct, for 
instance, then the absolute values of [Na/Fe] would be wrong, though their relative values (i.e. their differences) would remain valid.

Finally, let us consider the systematic errors due to the assumption of
local thermodynamic equilibrium (LTE) in the MOOG program. Non-LTE
(NLTE) calculations performed by Gratton et al. (\cite{GC99}) in
atmospheric conditions relevant to red giants show that the NLTE effect
on the Na\,\textsc{i}~$\lambda 6154.22$ line (one of the two lines used
in this paper) strongly depends on surface gravity and slightly
on effective temperature. This is illustrated in Fig.~\ref{Fig:NLTE}, which
reveals that the NLTE correction should be small in the $2.5<\log g<3$ range
characterizing our stars. Indeed, the average NLTE corrections are found to 
amount to 0.006--0.035~dex for the stars in NGC~2360 and of 0.032--0.048 dex
for NGC~2447 using Table~11 of Gratton et al. (quadratically interpolated in
$\log g$ and linearly in $T_{\rm eff}$, and assuming similar
corrections for both $\lambda 6154.22$ and $\lambda 6160.75$ lines of 
Na\,\textsc{i}). A similar calculation for iron\footnote{We
take the theoretical Fe\,\textsc{i} high excitation line ($EP=4.1$~eV) with
$W_\lambda=60$~m\AA~ (Gratton et al. \cite{GC99}) as representative 
of our data. Taking the low excitation line ($EP=0.0$~eV) or 
$W_\lambda=90$~m\AA~ would not change the results for the NLTE corrections.}
leads to NLTE corrections for Fe of 0.015--0.021 dex for  NGC~2360 and 
0.027--0.028 dex for NGC~2447. The resulting NLTE effects on [Na/Fe] thus 
ranges between -0.013 and 0.014 dex for NGC~2360 and between 0.005 and 0.020
for NGC~2447. Of course, these values are only approximate since they are 
derived from only one iron line assumed to be representative of the
55 lines observed for that 
element, but they do suggest that the errors brought by our LTE approximation
are much smaller than the above mentioned random error bars.

As a conclusion, the main abundance errors, besides the possible systematic
errors due to the oscillator strengths\footnote{The $\log gf$ values of the
Na\,\textsc{i} lines given by Boyarchuk et al. (\cite{B96}) are 0.05~dex higher
than those of Th\'evenin (\cite{T90}), but 0.04~dex lower than those of
Wiese et al. (\cite{W69}); this suggests that the systematic error on the
$\log gf$ values -- hence on the abundances -- cannot be much larger than 
0.05~dex. This is consistent with the error estimate of Th\'evenin.}, are due to the
equivalent width and continuum measurements, which amount up to 0.07~dex.
The error bars on the masses of each star, on the other hand, are
estimated by considering a 0.05 dex error on the age $\log t$ of the
clusters. The surface sodium overabundances as a function of stellar mass
are shown in Fig.~\ref{Fig:NaM} by rectangles taking into account the above
mentioned uncertainties.

\subsection{Predictions}
\label{Sect:Napredictions}

\begin{figure}
\center
\epsfig{file=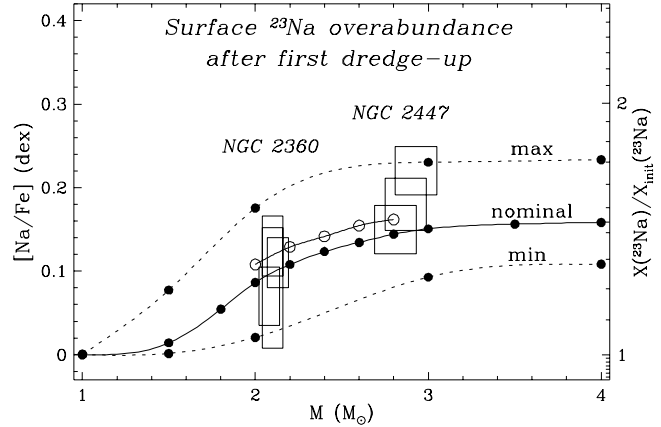,width=8.5cm}
\caption{\label{Fig:NaM}
         Na overabundance versus stellar mass. The rectangles represent the 
         abundance
         determined in our seven red giants with the estimated error bars.
         The theoretical curves are obtained at metallicity [Z=0.05]. The 
         continuous line joining the full dots is obtained with standard reaction
         rates and without overshooting, while that joining the open dots takes
         overshooting into account. The dotted lines show the predictions of
         a standard model (without overshooting) with extreme reaction rates
         still compatible with the uncertainties.
         }
\end{figure}

Sodium production during H-burning results from the transformation of
\chem{Ne}{22} into \chem{Na}{23} by proton capture. This reaction occurs
very efficiently at the temperatures characterizing the core of MS stars
(see Appendix~A of Mowlavi \cite{Mo99}). First dredge-up then mixes some
of the synthesized Na from the deep layers to the surface. This scenario
is confirmed by the observation of sodium overabundances at the surface of
many giants and supergiants (e.g. Luck \cite{L94}, Boyarchuk et al. \cite{B96},
Takeda \& Takada-Hidai \cite{TT94}).

The surface Na overabundance predicted by stellar model calculations
(without core overshooting) as a function of stellar mass is shown
in Fig.~\ref{Fig:NaM} by filled circles connected with solid line.
The models have a metallicity 0.05~dex above solar (which is the metallicity of
NGC~2447) and are followed from the pre-MS up to the completion of the 1DUP.
The stellar evolution code is the same as in Mowlavi
(\cite{Mo99}), except that the NACRE reaction rates (Arnould et al. \cite{AGJ99})
are used to follow the nucleosynthesis
and that the formalism of Canuto, Goldman \& Mazzitelli
(\cite{CGM96}) is used to describe the energy transport in convective
zones.
The surface sodium abundance is seen to be very sensitive to stellar
mass in the \mass{1.5-3} range. It increases from no Na enhancement below
\mass{\sim 1.5} to an overabundance of 0.15~dex at \mass{\sim 3}, and keeps this
value for stellar masses up to \mass{6} before increasing again with stellar mass
(see, e.g., Mowlavi \cite{Mo98}).

\begin{figure}
\center
\epsfig{file=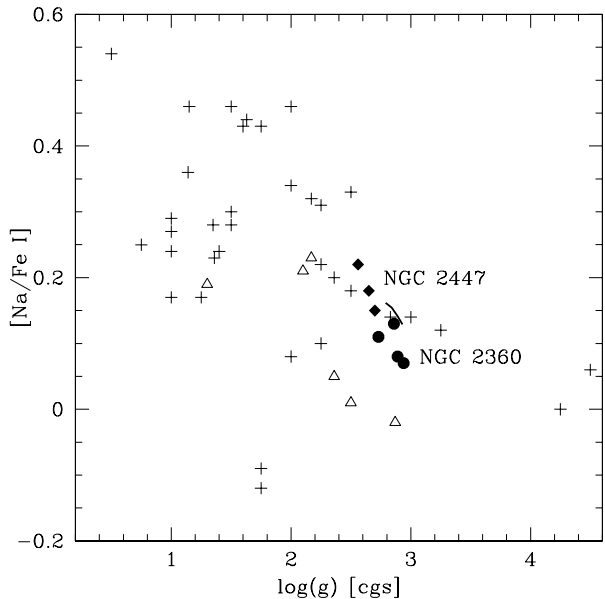,width=8.5cm}
\caption{\label{Fig:Nalogg}
Na overabundance versus $\log g$. The crosses represent yellow 
supergiants and giants in clusters measured by Luck \cite{L94}), the open
triangles represent red giants measured by Boyarchuk (\cite{B96}) and the full
symbols represent the seven red giants of this study (full dots: NGC 2360,
full diamonds: NGC 2447). The two crosses at high $\log g$
are dwarfs, while the two supergiants with a negative [Na/Fe] value are among
the coolest in Luck's sample ($T_{\rm eff} = 4000$~K). The short solid line on the right of our observations corresponds to predictions of clump stars of 
2.2 to \mass{2.8} computed with core overshooting.}
\end{figure}

The sensitivity of those predictions to core overshooting, convection
prescription and stellar metallicity is explored by computing extra models from
the pre-MS to the 1DUP.
None of those parameters, however, turns out to have a significant impact on
the surface Na abundance.  Models with core overshooting (with an
extra-mixing extent of 0.20 times the pressure scale at the core
boundary) predict a 0.02~dex enhancement (open circles connected with
solid line in Fig.~\ref{Fig:NaM}) compared to predictions without
core overshooting.
Increasing the metallicity by 0.17~dex does not change the surface [Na/Fe] prediction
after 1DUP by more than 0.01~dex.
And using the mixing length theory (with a
mixing length of 1.5 times the pressure scale height) instead of the
Canuto, Goldmann \& Mazzitelli formalism does not change the surface
abundance predictions within 0.001~dex.

Let us now explore the uncertainties linked to nuclear reaction
rates. Both the \reac{Ne}{22}{p}{\gamma}{Na}{23} and p-capture reactions on
\chem{Na}{23} are still subject to large uncertainties (Arnould et al.
\cite{AGJ99}). In order to assess their impact on our surface Na abundance
predictions, several models are recomputed from the pre-MS up to the
1DUP with the upper/lower limits for the rates provided by the NACRE compilation
(cf. Arnould et al.), as appropriate to minimize/maximize \chem{Na}{23}
production.
The results in the `minimal' and `maximal' \chem{Na}{23} production cases
are shown in filled
circles connected with dotted lines in Fig.~\ref{Fig:NaM}. They reveal
a variation in the surface Na abundance predictions of up to 0.08~dex relative to
the `nominal' case where the recommended NACRE rates are used. Nuclear 
reaction rate uncertainties thus dominate the uncertainties associated with 
stellar metallicity and convection prescriptions for sodium predictions in red 
giants.

\subsection{Discussion}
\label{Sect:Nadiscussion}

Figure~\ref{Fig:NaM} shows a
very good agreement between our Na abundance predictions in \mass{2} model stars
and those observed in NGC~2360. The nominal predictions of the \mass{3} model
star, on the other hand, seem a little too low compared to the abundances
observed in NGC~2447. The predictions in the maximal case of \chem{Na}{23} production
would fit the highest Na abundance measured among
the three stars observed in NGC~2447. Those predictions,
however, would not be compatible with the Na abundances
measured in the \mass{2} red giants of NGC~2360. Figure~\ref{Fig:NaM}
thus suggests that the nominal Ne--Na reaction rates should not be
too much altered, if at all. The solution to the discrepancy between our
observed Na abundance observations in NGC2447 and predictions should be found
in other(s) mechanism(s) such as, possibly, meridional mixing induced by
stellar rotation. Further theoretical and observational investigations should
be performed before being able to draw a firm conclusion.

Finally, let us mention that the positive sodium abundance -- stellar mass
correlation translates, at a given effective temperature into a sodium abundance
-- surface gravity anti-correlation (or sodium abundance -- luminosity
correlation). This is well known in the literature, and shown in
Fig.~\ref{Fig:Nalogg} where our data are displayed together with those of
Luck (\cite{L94}) and Boyarchuck et al. (\cite{B96}). The dependence on
effective temperature is small (Luck \cite{L94}).
The computation of our 2.2 to \mass{2.8} models with core overshooting is
carried on up to the clump in the core helium burning phase. The clump is
defined as the point where the stellar luminosity (gravity) reaches its minimum
(maximum) value after core helium ignition. The sodium abundance (which is not
altered between the 1DUP and the clump) predictions for those clump models are
shown by a solid line in Fig. 6.

\section{Oxygen}
\label{Sect:oxygen}

\subsection{Abundances}
\label{Sect:Oabundances}

The O abundance is determined from the 6300.31\AA~ line of O\,\textsc{i}, which
is available in our spectra with a S/N ratio above 100 only for the stars of
NGC~2447 and for one of the stars in NGC 2360.
This line is slightly blended with a line of Sc\,\textsc{ii} at 6300.68\AA, but
both lines are fairly well separated in our spectra
(see Fig.~\ref{Fig:O-Sc}). Fortunately, no significant telluric line is
spoiling the O and Sc lines, though some are present 1 \AA~ away or more.
Using the equivalent width of the 6300.31\AA~ line,
the O abundance is found to be slightly deficient with respect to iron,
with [O/Fe]$ = -0.1$ to $-0.3$~dex (see Table~\ref{Tab:abundances}).

The O\,\textsc{i} line at 6300.31\AA, however, may be blended with
another line at 6300.33\AA~ from Ni\,\textsc{i}.
The $\log gf$ of the Ni\,\textsc{i} line is still poorly
known. If we take the only data available in the recent literature,
i.e. $\log gf = -1.737$ and $EP = 4.266$~eV proposed  by Kurucz (\cite{K94}),
and if we assume [Ni/Fe]=0.00, then the oxygen
abundance predicted by synthetic spectra reaches
-0.6~dex on average (cf. [O/Fe]$^*$ in Table~\ref{Tab:abundances}).
A smaller value of $\log gf$ is however suggested from the analysis of the
solar spectrum. Lambert (\cite{L78}) estimates the width $W_\lambda$ of the
Ni\,\textsc{i} $\lambda 6300.33$~ \AA~
line in the solar spectrum to be below 0.1 -- 0.5 m\AA. The oscillator strength
derived from the solar Ni abundance would be ten times smaller than Kurucz's 
even with $W_\lambda=0.5$~m\AA. The true $\log gf$ of this Ni\,\textsc{i} line 
is thus very probably smaller than -2.74 dex.
Therefore, the blend cannot affect the oxygen abundance determination by more
than a few hundredths of dex.

\subsection{Predictions}
\label{Sect:Opredictions}

\begin{figure}
\center
\epsfig{file=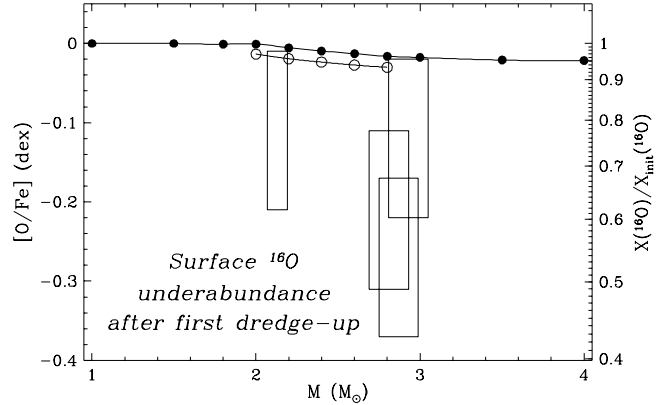,width=8.5cm}
\caption{\label{Fig:OM}
         Same as Fig.~\ref{Fig:NaM}, but for oxygen.
}
\end{figure}

The surface O underabundance predictions after the first dredge-up are shown in
Fig.~\ref{Fig:OM} as a function of stellar mass. Models with (open circles)
and without (filled circles) core overshooting during MS both predict very little
variation of the surface O abundance in red giants (depletion by less than 0.04
dex). This results from the fact
that only the deepest stellar layers are affected by the N--O cycle of H-burning.
As a result, the dilution factor of the envelope's oxygen is rather small when the
envelope extends down to the inner regions.

\subsection{Discussion}
\label{Sect:Odiscussion}

The predicted surface O underabundances fall much below the observed ones (by
0.1 -- 0.3 dex if Ni is not taken into account and by more than 0.5 dex with Ni).
This discrepancy is unexplained so far.

Such a deficiency of O in red giants has already been reported in the literature
(e.g. Luck \cite{L94}).
Interestingly, Venn (\cite{V99}) argues that the Sun presents an overabundance
of oxygen of about 0.3 dex relative to the galactic A supergiants, to B stars
in the solar neighborhood and in Orion, and to nebular abundances in the Orion
nebula. The conclusion that the solar oxygen abundance is higher than what
should be expected by galactic chemical evolution is
also reached, at least qualitatively, from the sample of G dwarfs of
Edvardsson et al. (\cite{E93}), which show an average oxygen abundance of
[O/Fe]$\sim$-0.1 at [Fe/H]$>$-0.1.
If this empirical fact is confirmed, it would explain in a simple way the
oxygen underabundances found in our giants, at least if we neglect the
contribution of the Ni\,\textsc{i} line in the synthetic spectra.

\begin{figure}
\epsfig{figure=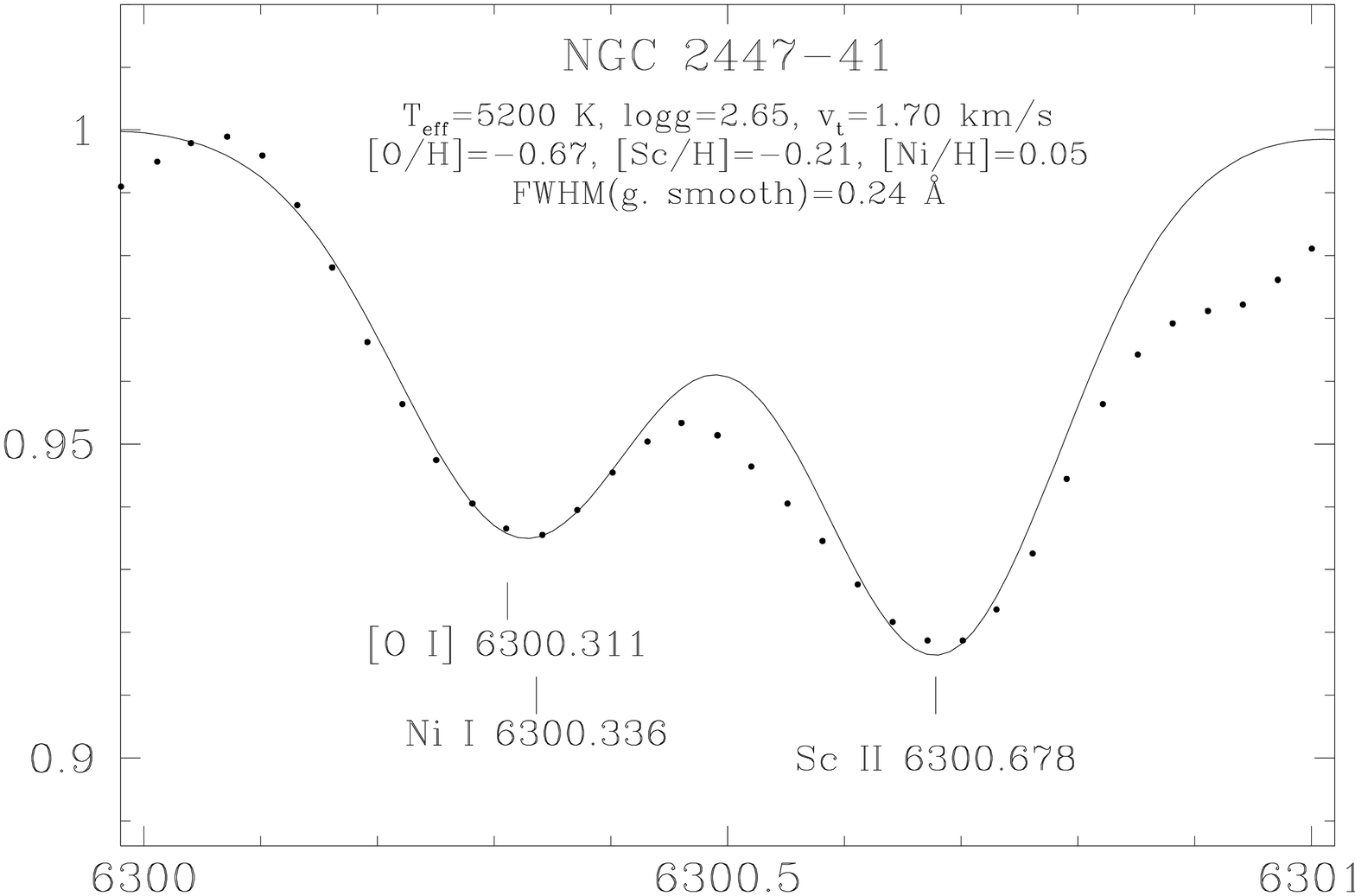,width=8.5cm}
\caption{\label{Fig:O-Sc}
         Measured spectrum of the star NGC 2447-41 (dots), showing how the 
         [O\,\textsc{i}] line used for the abundance determination is slightly blended
         with an Sc\,\textsc{ii} line. The continuous line is the synthetic spectrum
         computed with MOOG, including the [O\,\textsc{i}] 6300, Sc\,\textsc{ii} and
         Ni\,\textsc{i} lines. The abundances adopted for the fit are given on the plot.
         The measured spectrum has been slightly smoothed using the Fourier method,
         with a window of 0.15~\AA.}
\end{figure}

\section{Conclusions}
\label{Sect:conclusions}

The abundances of about fifteen elements in seven red giants of
NGC~2360 and NGC~2447 are derived with a global uncertainty of about $\pm 0.1$~dex
(essentially due to photon noise, normalization of the spectra and uncertainties
related to atomic line parameters). From those abundances, the following
conclusions are drawn.

The metallicity [$M$/H] of the two open clusters is directly derived from their
iron-group elements. It amounts to 0.10 for NGC 2360 and 0.05 for NGC 2447.

The sodium abundance reveals a positive correlation with stellar mass in the
\mass{2-3} range, as expected from model predictions. Quantitatively, the surface
\chem{Na}{23} abundances are in good agreement with those predicted after 1DUP for
\mass{2} stars (i.e. for stars in NGC~2360), but not at \mass{3} (NGC~2447). The
disagreement in NGC~2447 reaches 0.08~dex, which we consider marginally 
significant given our estimated uncertainties on the Na abundances. The excess 
of Na observed in two of our stars of NGC~2447, if confirmed, cannot be 
explained by current stellar models.
Other physical mechanisms, such as stellar rotation, should be investigated in
the future. More observations in open clusters of various ages (and thus 
turn-off masses) are also required to increase the small number statistic of 
this study.

The oxygen abundance is very deficient (by 0.2 to 0.5~dex depending on
the true $gf$ value of the Ni\,\textsc{i} $\lambda 6300.34$ line) relative to 
iron in all our four red
giants for which the oxygen abundance could be measured. This is not compatible
with stellar model calculations which predict an oxygen deficiency by at most
0.03~dex after the first dredge-up.
The high deficiencies can be partly explained if we assume that the Sun is
slightly oxygen-rich relative to most stars in its neighborhood, as
suggested by several authors. If not, then a new mechanism
in stellar models should be thought of to account for this discrepancy.

\begin{acknowledgements}
We thank Dr. Pierre Dubath for his help with the INTER-TACOS software used for
the reduction of echelle spectra, and Dr. Laura Fullton for providing us
with Dr. Sneden's ftp address from which we downloaded the MOOG code.
PN thanks Dr. Georges Meynet for a fruitful discussion. This work was
supported in part by the Swiss National Science Foundation.
\end{acknowledgements}

\end{document}